 \documentclass[final,5p,times,twocolumn]{elsarticle}

\usepackage{amssymb}
\journal{Nuclear Physics B}

\begin{document}

\begin{frontmatter}

%% Title, authors and addresses

%% use the tnoteref command within \title for footnotes;
%% use the tnotetext command for theassociated footnote;
%% use the fnref command within \author or \address for footnotes;
%% use the fntext command for theassociated footnote;
%% use the corref command within \author for corresponding author footnotes;
%% use the cortext command for theassociated footnote;
%% use the ead command for the email address,
%% and the form \ead[url] for the home page:
%% \title{Title\tnoteref{label1}}
%% \tnotetext[label1]{}
%% \author{Name\corref{cor1}\fnref{label2}}
%% \ead{email address}
%% \ead[url]{home page}
%% \fntext[label2]{}
%% \cortext[cor1]{}
%% \address{Address\fnref{label3}}
%% \fntext[label3]{}

\title{Silencing `noisy' 2DEG wafers }

%% use optional labels to link authors explicitly to addresses:
%% \author[label1,label2]{}
%% \address[label1]{}
%% \address[label2]{}

\author{L. Gaudreau$^{a,b}$, A. Kam$^a$, P. Zawadzki$^a$, G. Granger$^a$,  S. A. Studenikin$^a$, J. Kycia$^c$, J. Mason$^c$,  and A. S. Sachrajda$^a$ }

\address{$^{a}$Institute for Microstructural Sciences, National Research Council, Ottawa, Canada, K1A 0R6\\
$^{b}$Physics Department, University of Sherbrooke, Quebec, Canada, J1K 2R1\\
$^{c}$Department of Physics and Astronomy, University of Waterloo, Ontario, Canada, N2L 3G1}

\begin{abstract}
Telegraphic noise is one of the most significant problems that arises when making sensitive measurements with lateral electrostatic devices. In this paper we demonstrate that a wafer which had only produced devices with significant telegraph noise problems can be made to produce `quiet' devices if a thin insulator layer is placed between the gates and the GaAs/AlGaAs heterostructure. A slow drift in the resulting devices is attributed to the trapping of charges within the specific insulator used. This charge can be manipulated, leading to strategies for stabilizing the device.

\end{abstract}

\begin{keyword}
%% keywords here, in the form: keyword \sep keyword
Telegraph noise; Schottky barrier; Leakage current

%% PACS codes here, in the form: \PACS code \sep code

%% MSC codes here, in the form: \MSC code \sep code
%% or \MSC[2008] code \sep code (2000 is the default)

\end{keyword}

\end{frontmatter}

%% \linenumbers

%% main text
\section{Introduction}
\label{Introduction}

Many sensitive experiments involving lateral devices are negatively affected by switching noise. These devices are usually defined in the two dimensional electron gas (2DEG) within GaAs/AlGaAs heterostructures by means of electrostatic gates. Recently we showed that the origin of the noise lay in an interrupted leakage current from the gates used to define the device\cite{Pioro-Ladriere2005}. This has led to several strategies to mitigate the noise. These include the use of `bias cooling'\cite{Pioro-Ladriere2005} which involves the application of a positive voltage to the gates during cooldown and `global gating'\cite{Buizert2008} which involves the application of a negative voltage to a gate covering the whole device. Both of these techniques increase the effective tunnel barrier at the metal-semiconductor interface and thus lower the leakage current and hence the noise. In this paper, we show that by eliminating this interface altogether we are able to convert a 'noisy' wafer, i.e. one which has a history of generating only noisy devices, into one where the telegraph noise is eliminated. In addition to providing a route to making useful devices from problematic wafers, these experiments provide further evidence for our model for the origin of the noise.

\section{Experiment}
\label{Experiment}

\begin{figure}
\begin{center}
\includegraphics*[scale=0.9]{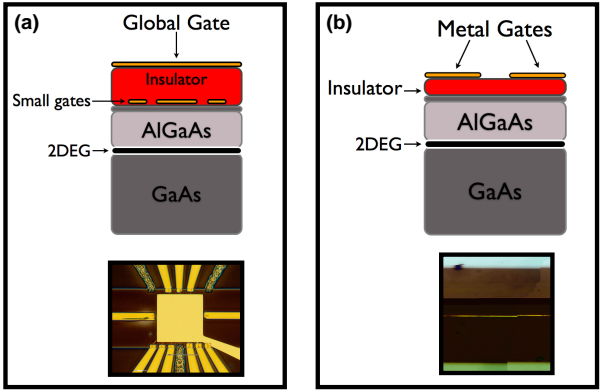}
\caption{(a) Schematic representation of the global gate deposition process. The Schottky gates are covered with a 40nm thick layer of calixarene  and a AuTi global gate is deposited on top. The calixarene serves as an insulator between small and global gates. Optical microscope photograph of a device fabricated with this technique. (b) Schematic representation of the floating gates technique. The Schottky gates are deposited on top of the calixarene layer to remove the metal-semiconductor interface through which the leakage current leading to telegraphic noise occurs. Optical microscope photograph of a QPC where the small gates (yellow) can be seen on top of the calixarene layer (dark).}
\label{fig:1}
\end{center}
\end{figure}

While the origin of the telegraph noise lies in an interrupted leakage current, it is an empirical fact that some wafers are much more `noisy' than others.  For the experiments in this paper we used our `noisiest' wafer which historically had only produced devices with significant telegraph noise problems. For the global gate experiments, a large global gate was fabricated on top of three 40 nm layers of calixarene. This number of layers was found to be necessary to prevent shorting between gates due to pinholes in the calixarene. In the second 'floating gate' experiments, QPCs and quantum dots were fabricated on top of a 40nm single layer of calixarene in separate fabrication runs.  The layer structures are shown schematically in figure \ref{fig:1}.  A photograph of one of the QPCs on top of the calixarene, as well as a global gate device are also shown.  Calixarene is a class of marcomolecules with relatively low molecular weight which is also a high resolution negative electron beam resist. Moreover, due to its robustness upon exposure and development, it can be employed as spacer dielectric layer between our fabricated mesa and the fine gate structure used to form our quantum dots. This layer was achieved by spinning calixarene dissolved in cholorbenzene at 2000 rpm for 30 seconds and curing at 180$^\circ$C for 5 minutes. This gave a 40nm thick layer.  To produce the required gate on top of the calixarene, the sample was then lithographically patterned by an electron beam and developed in o-xylene. The cross-linked calixarene formed a protective layer over the mesa and acted as a template on which the fine gates were patterned and suspended over the mesa.

Noise tests were made by monitoring transport through the QPC which was placed at a sensitive charge detection voltage. Measurements were made at 1.4 K using standard low noise transport measurement techniques. Identical results were obtained on both the QPC devices and the QPCs formed within the quantum dot devices.

\section{Results and discussion}
\label{Results and discussion}

\begin{figure}
\begin{center}
%\begin{picture}(5,10)(0,0)

%\includegraphics[width=8.0cm, keepaspectratio=true]{fig2}

\includegraphics*[scale=3]{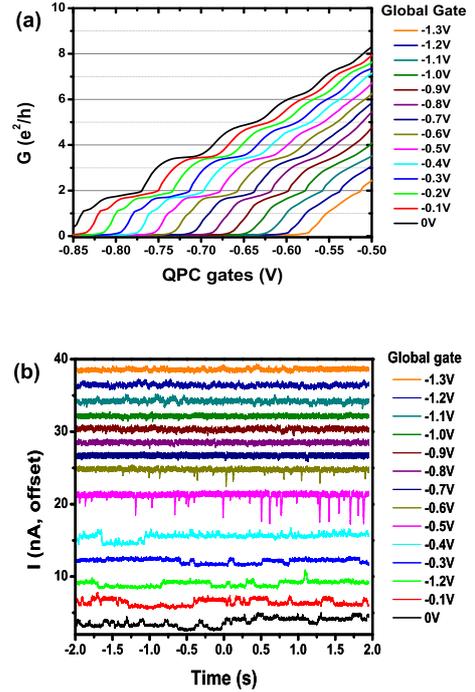}

%\end{picture}
\end{center}
\caption{(a) QPC conductance traces taken at different global gate voltages. (b) Time traces of the QPC set at the high sensitivity operation point for different global gate voltages. At 0V on the global gate, telegraphic noise is measured (black curve). As a negative voltage is applied on the global gate, the noise is reduced.}
\label{fig:2}
\end{figure}

Figure \ref{fig:2}(a) shows QPC traces from a QPC under a global gate as the global gate voltage is varied. The QPC traces can be seen to shift to less negative voltage values as a negative voltage is applied to the global gate. It is interesting to note that in these traces the width and electron density in the QPC are different for the same QPC plateaus. This was found, in particular, in some devices to dramatically affect the quality of the 0.7 feature\cite{Thomas1996}. More details of this will be published elsewhere. Figure \ref{fig:2}(b) shows time traces for the different global gate voltages. The QPC voltages are set at the point at which they are highly sensitive to the local electrostatic environment. It can be seen that at the more negative global gate voltages, the noise is dramatically reduced. We find that there are some resonant global gate voltage values at which the noise returns. Details of these experiments will be published elsewhere. The reason why the global gate technique works is similar to why bias cooling works, i.e. the voltage values at the metal-semiconductor interface is reduced and therefore the tunnelling barrier is increased, hence the leakage current, an essential element in producing telegraph noise, is reduced.  A limitation of the global gate technique is, however, that the same voltage adjustment is made to all device gates at the metal-semiconductor interface which for complicated quantum dot circuits may not be optimum. It is therefore beneficial to introduce approaches which eliminate the leakage current without the need for global gates or bias cooling.

\begin{figure}
\begin{center}
\includegraphics*[scale=1.8]{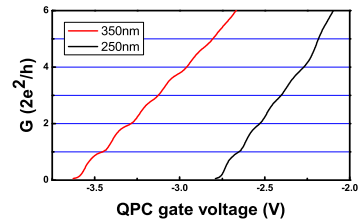}
\caption{ Conductance traces of two QPCs with different widths showing the pinch-off values at more negative voltages due to the increased distance between the gates and the 2DEG.}
\end{center}
\label{fig:3}
\end{figure}

Figure \ref{fig:3} plots the conductance traces of two of the QPCs fabricated using the floating gate approach. The QPCs are defined wider than usual due to the extra 40nm distance between the 2DEG and the gates. For the same reason, the applied gate voltages are much more negative over the QPC operation range.  It is apparent, however, that standard QPC behaviour is observed. 

\begin{figure}
\begin{center}
\includegraphics*[scale=0.75]{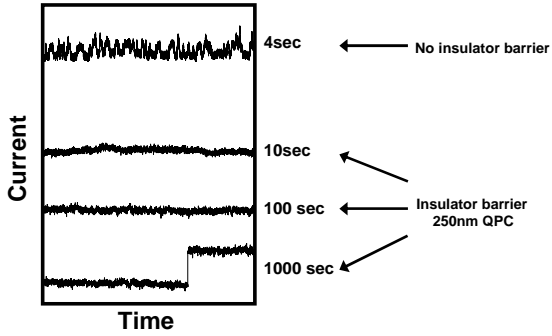}
\caption{Time traces taken at the high sensitivity point for two QPCs: one fabricated with the insulating layer (bottom three curves) and another one without it (top curve), both on the same wafer. The top curve shows telegraphic noise, while in the bottom three, only one event is detected over a large time scale, which is related to charge trapping in the calixarene layer.}
\label{fig:4}
\end{center}
\end{figure}

Figure \ref{fig:4} plots several traces of the QPC conductance for different time intervals. For these traces the QPC is placed at a sensitive charge detection point. The top trace is a typical trace from a QPC fabricated on this wafer with no insulator. A large amount of telegraph noise can be clearly seen.  The bottom three traces are taken from the 250nm QPC for different time intervals. It is clear that no telegraph noise is visible. A similar lack of noise is obtained on all of the QPCs and quantum dots studied in devices fabricated with the insulator spacer. This confirms the importance of the semiconductor-metal interface and the role of the leakage current in generating the switching noise.

\begin{figure}
\begin{center}
\includegraphics*[scale=1.02]{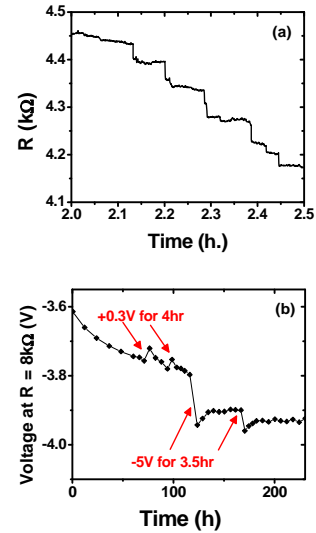}
\caption{(a) QPC resistance as a function of time at a fixed gate voltage. Discrete steps are observed and are attributed to charge dynamics within the calixarene layer. The relaxation time for reaching an equilibrium value is close to two weeks. (b) Gate voltage required to keep the QPC resistance at 8 k$\Omega$ plotted vs. time. The deviations from normal relaxation correspond to the application of additional voltages to the QPC gates, as explained in the text.}
\label{fig:5}
\end{center}
\end{figure}

The bottom trace in figure \ref{fig:4} contains a single jump in the QPC conductance. This is not, however, a telegraphic noise jump. We have found that all devices grown with the  calixarene spacer 'relax' in a digital fashion when a voltage is applied to the QPC gates. The relaxation always takes place in the direction of decreased resistance. A section of a much longer trace is shown in figure \ref{fig:5}(a).  In this trace about ten jumps can be seen over 30 minutes. The relaxation time for reaching an equilibrium value with no additional strategy is close to two weeks.

It should be noted that because calixarene is a conjugated molecule it can trap charges and any conductivity in calixarene likely involves the hopping of charge over defects\cite{BenChaabane1997}. It is therefore reasonable to assume that by applying a negative voltage to the gates defining a QPC, negative charge is gradually pushed away from the region close to the gates. Since this charge contributes to the electrostatic environment within the QPC, the current in the QPC increases (bottom trace in figure \ref{fig:4}) or equivalently, the resistance decreases (figure \ref{fig:5}(a)).  The digital nature of the relaxation suggests that reconfigurations involving multiple charges are taking place.

To test these ideas and to identify strategies for reaching faster relaxations we adjusted the gate voltages on the QPCs and studied the response. An example is given in figure \ref{fig:5}(b). In these measurements the gate voltage applied to the QPC gates is adjusted to maintain the resistance at a constant value, in this case 8 k$\Omega$. Initially, on application of a QPC gate voltage the system slowly relaxes with the voltage becoming more negative indicating the gradual removal of negative charge in the calixarene from the vicinity of the QPC gates. At a certain time, a positive voltage is applied to the QPC gates and after a 4 h period, the measurements are recommenced. The procedure results in a less negative voltage to keep the QPC at 8k$\Omega$, confirming that negative charge within the calixarene layer has been attracted to the vicinity of the QPC gates. The process is reproducible as seen in the figure. On applying a negative charge for several hours to the gates, the opposite and consistent behavior happens. After a short charge distribution relaxation, the system now stabilizes at a constant voltage. The gate voltage to maintain 8 k$\Omega$ is now more negative, suggesting the removal of negative charge within the calixarene layer. However, the charge distribution for stability is not unique since repeating the procedure again creates a stable device, but at a different gate voltage value suggesting a different charge distribution.

\section{Conclusions}
\label{Conclusions}

We have confirmed the advantages of a global gate in reducing noise. We have further shown that a noisy GaAs/AlGaAs wafer can be made free of telegraph noise by use of a floating gate technique. These two experiments confirm our understanding of the noise as being due to an interrupted leakage current at the metal-semiconductor interface. In the floating gate approach we find a very slow drift behaviour consistent with trapped charge in the calixarene. This charge can be manipulated by applying large positive or negative voltages to the gates. Such operations can stabilize the device, but the charge distribution at which this is achieved is not unique. Although stable devices are the ultimate aim of this project, clearly it would be optimum not to have the drift in the first place. Other insulators are being investigated for this purpose.

\section{Acknowledgements}
\label{Acknowledgements}
We thank J. Gupta and Z. Wasilewski for useful discussions. A.S.S and J.K acknowledge the support of NSERC and the Quantumworks network. A.S.S. also acknowledges support from CIFAR.

%% The Appendices part is started with the command \appendix;
%% appendix sections are then done as normal sections
%% \appendix

%% \section{}
%% \label{}

%%******************************************************************

%\bibliographystyle{physicaE}
%\bibliography{Telegraphicnoise_EP2DS2009}

%%******************************************************************

%%*****************************************************************
\end{document}